\documentclass[conference,compsoc]{IEEEtran}
\ifCLASSOPTIONcompsoc
  \usepackage[nocompress]{cite}
\else
  \usepackage{cite}
\fi
\ifCLASSINFOpdf
  \usepackage[pdftex]{graphicx}
  \DeclareGraphicsExtensions{.pdf,.jpeg,.png}
\else
\fi
\usepackage{amsmath}
\usepackage{algorithmic}
\usepackage{algorithm2e}

\usepackage{array}
\usepackage{url}

\begin{document}
%
%
\title{Enabling Community Health Care with Microservices}
%
\author{\IEEEauthorblockN{Richard Hill}
\IEEEauthorblockA{School of Computing and Engineering\\
University of Huddersfield\\
United Kingdom\\
Email: r.hill@hud.ac.uk}
\and
\IEEEauthorblockN{Dharmendra Shadija and Mo Rezai}
\IEEEauthorblockA{Department of Computing\\
Sheffield Hallam University\\
United Kingdom\\
Email: \{d.shadija\},\{m.j.rezai\}@shu.ac.uk}
}
%

%

%
%
\maketitle
%
\begin{abstract}
Microservice architectures (MA) are composed of loosely coupled, coarse-grained services that emphasise resilience and autonomy, enabling more scalable applications to be developed. Such architectures are more tolerant of changing demands from users and enterprises, in response to emerging technologies and their associated influences upon human interaction and behaviour. This article looks at microservices in the Internet of Things (IoT) through the lens of agency, and using an example in the community health care domain explores how a complex application scenario (both in terms of software and hardware interactions) might be modelled.\\
\end{abstract}
%
Microservices, health care, Internet of Things (IoT), Agent Unified Modeling Language (AUML)
%
%
%
\IEEEpeerreviewmaketitle

%




%
\section{Introduction}
The delivery of healthcare services within a community setting is a fundamental part of an effective care regime that enables recipients to recuperate or cope with personal challenges in their home environment. Within the United Kingdom, government policy has resulted in changes to the way that health and social care services are delivered\cite{beer2003}, with an emphasis upon developing a competitive market place by engaging more private sector providers.
Community health care is typically a complex set of services that incorporates a number of different stakeholders, each with their own agendas and objectives. Primarily, these services are to maintain or enhance the quality of life of a recipient, in order that they can retain their independence.
There is also the consideration of costs; health care services have to operate within budgetary constraints and the managers of such services must deliver their responsibilities in an efficient and timely manner.
A large proportion of the community health care activities are focused upon communication and coordination between various agencies. Care delivery requires people and resources to be mobilised and deployed in the home of the care recipient, together with the associated communication of any changes to the individuals' care requirements before and after care delivery.

In addition, the care provider must manage the delivery of its services and report back for the purposes of quality assurance.

As such, the provision of community health care services could be represented by a multiagent system that brokers and negotiates objectives between the different agencies, enabling care data to be exchanged securely and resources to be delivered efficiently. The system proposed by Huang et al\cite{huang1995} used an agent-oriented architecture to not only support traditional care services, but also additional roles that had not been formally part of the care delivery system before. 

The inclusion of informal care delivery (such as a local Warden or neighbour who provide social contact and are often the first agents to raise alarms for emergency care), means that any resulting system would have a much richer picture of the care requirements of an individual recipient.

This would then enable the package of care to be tailored to the individual needs of a recipient, whilst also facilitating a more agile response to changing conditions. This particular aspect is most pertinent when delivering palliative or end-of-life care, when the condition of the care recipient can deteriorate too quickly for traditional care systems to be able to respond in a timely fashion, jeopardising the quality of life objective that the care system is attempting to address.

This article examines the community health care domain using a service oriented architecture that is modelled as a multiagent system, and is organised as follows. First, we briefly review the need for service oriented architectures, and in particular why a Microservice Architecture (MA) might be suitable for the community health care domain. Second, we examine some specific scenarios that illustrate the complexities of a solution architecture, and consider the relvance of an agent oriented approach to modelling. We then propose a microservice based IoT architecture, and experiment by way of simulation to explore the data transfer requirements.
Finally we discuss the limitations and opportunities afforded by an agent managed approach to microservices in the community health care domain.
\section{Software engineering}
%
The discipline of software engineering has been established for some time now\cite{wirth2008}, with a variety of approaches, techniques and tools being created to assist software application designers and developers. We have considered the domain of community health care through the lens of agency and service based architectures; in particular, the use of a coarse-grained Microservice Architecture. For a more comprehensive discussion of MA, see Shadija et al\cite{shadija2017}.
\subsection{Software architecture}
Application logic was represented by Jackson Structured Programming (JSP)\cite{jackson1975}, which encouraged the maintenance of a library of cohesive subroutines, along with a clear separation between data and program structures. Such an approach promotes modularity and reuse of program code. Object Orientation (OO) was a further development\cite{dahl1972}, where service-like objects were invoked by other objects\cite{snyder1991}, providing abstraction away from the often complex internal logic of an application\cite{walker2007,berners,alonso2004,weerawarana2005}. Each object encapsulates data relevant to the object.\cite{snyder1991}. 

Modularity is a primary objective of OO, to enable maximum reuse of code, at a low level of granularity.
One of the issues of fine-grained granularity is the increased dependency between objects on code that is reused. Snyder argues that a component-based approach (that is packaging objects together as a component) would facilitate a greater level of software development productivity as the services would be coarser-grained and therefore more representative of the business logic\cite{snyder1991}.
\subsection{Service Oriented Architecture}
The paradigm of Service Oriented Architecture (SOA)\cite{mackenzie2006} provides further encapsulation by aligning an interface to a number of objects, and then to a discrete business function, which is more coarse-grained than pure componentisation\cite{alahmari2010}. Like OO architectures, services exchange messages between each other to consume other services at runtime through late binding\cite{mos2013}. To promote reuse and interoperability, industry standard protocols such as SOAP are utilised.
\subsection{Agent-Oriented Software Engineering}
Agent-Oriented Software Engineering (AOSE) is a variation of traditional software engineering approaches. Whilst OO development attempts to simplify software application design by narrowing the gap between program code and the `real world' through object representations, the essential characteristics of passive objects do not support the dynamic and proactive abilities possessed by real-world agents. AOSE embraces agency and faithfully represents the more decentralised systems and their associated interactions, thus making this approach more suitable for complex applications.

Such systems (for instance community health care delivery) require agency in order for various stakeholders to be able to take the initiative, negotiate and broker actions and data in a dynamic, heterogeneous environment. The increased capabilities of agents permits complex organisational workflows to be represented, whilst also enabling the mapping of existing organisational models to agent representations, in order to represent inter-dependencies and complex interactions\cite{luck2004}.
\subsection{Microservice Architecture}
Microservice Architectures (MA) utilise services that address a single business capability, with a clearly defined interface\cite{lewis2014}. They are cohesive and loosely coupled to other microservices, to perform a larger business function. This architecture is enabled by each microservice owning a data and class model, which facilitates resilience in the eventual application. Using the Domain Driven Design (DDD) approach\cite{evans2003,evans2015}, Evans posits that a bounded context should inform the decomposition of program code, and consequently its subsequent reuse.

We shall now consider example scenarios relevant to the delivery of community health care.
\section{Community health care scenarios}
\begin{figure}[!t]
\centering
\includegraphics[width=3.15in]{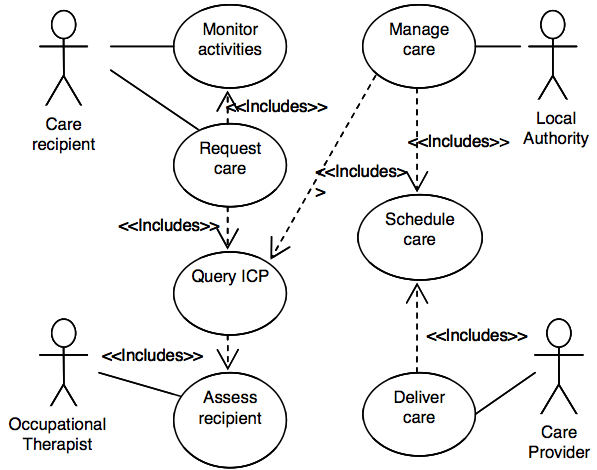}
\caption{Overall community health care use case model.}
\label{fig:healthcareoverview}
\end{figure}

In the context of community health care delivery, Beer et al\cite{beer2002}, describe five separate scenarios that a system would need to represent and support as follows:
\begin{enumerate}
\item An Individual Care Plan (ICP) is a living document that specifies the care services that an individual should receive. This document is maintained in accordance with the changing needs of the care recipient;
\item The delivery of care services to a care recipient in order that their quality of life is improved;
\item In addition to (2), the delivery of routine care to support the independence of the care recipient;
\item Delivery of emergency care when required in a timely fashion;
\item Management of the myriad services and agents in order that quality is assured, interacting with the ICP as necessary.
\end{enumerate}
Figure \ref{fig:healthcareoverview} describes the overall community health care scenario in Agent Unified Modelling Language\cite{odell2001}, including stakeholders and dependencies between use cases\cite{hill2005}.
\subsection{Individual Care Plan management}
Central to the provision of care is the Individual Care Plan (ICP), which contains information about the care recipient in the form of medical assessments made of their capabilities. A package of care services is then assembled to address the needs identified by the assessments. Traditionally, ICPs contained information that was produced exclusively from assessments by Occupational Therapists or other medical professional staff. Such assessments are costly to produce and as a result there are limited resources to update and maintain assessments. 

However, technologies that are being developed for the Internet of Things (IoT)\cite{ikram2015},\cite{bessis2013} are providing new opportunities to remotely sense, collate, analyse and monitor data about care recipients whilst they continue to live in their home environments. IoT developments, particularly in the context of the Industrial Internet of Things (IIoT), are enabling new methods of analysis and reporting for proactive monitoring\cite{hill2017}  in environments that utilise remote sensors. There are two key benefits of remote sensing. 

First, the recipient is being continuously monitored which enables the ICP to be more responsive to the needs of the recipient. Second, the data collected pertains to the activities of the recipient in their home environment, which also enables the package of care services to be tailored to their specific needs. Whilst the medical needs of two individual care recipients might be identical, the environmental situation of the recipients may necessitate a change in how the care services are delivered. This contrasts with more established systems that would provide the same service to different care recipients, irrespective of their home environments.
%
%
\subsection{Improving quality of life}
Improving (or at least reducing the rate of decline) of quality of life is a primary concern for community health care delivery. Any failure to address this aspect results in at least some discomfort for the recipient, but depending upon the nature of the care required, can be life-threatening. Since the potential outcome is quite severe, upon detection of an event or situation that threatens quality of life, the care managers typically flood the situation with resource until stability is regained. This is a rather wasteful approach to the management of scarce resources, especially since evaluations of such situations reveal that they could have been prevented in the first place had more effective coordination and response systems been in place.
Two factors that directly contribute to improving the quality of life are a) effective social interaction between the care recipient and care providers, and b) the delivery of opportunities to engage the care recipient in new leisure experiences according to their personal preferences. One advantage of enhanced social interaction between all of the stakeholders in community health care provision is the potential for improved communication of pertinent, timely information, that can be used to improve the quality of care delivered. 
%
%
\subsection{Routine care}
\begin{figure}[!t]
\centering
\includegraphics[width=3.15in]{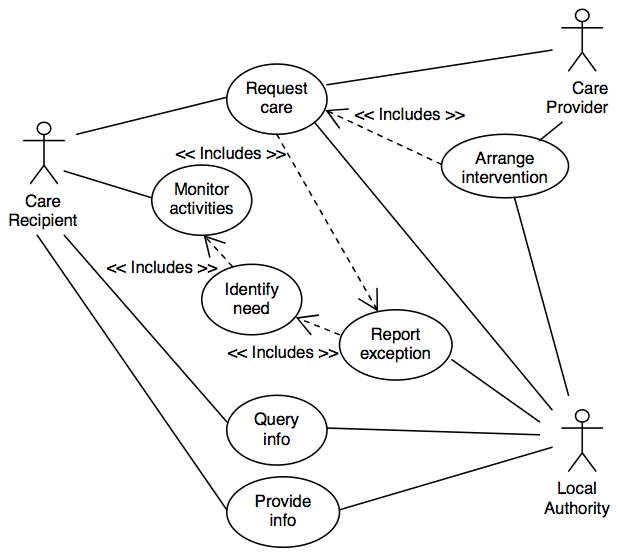}
\caption{Routine care use case model.}
\label{fig:routinecare}
\end{figure}

An important part of care provision is that of routine care (Figure \ref{fig:routinecare}); assistance with daily functions (eating, washing, regular medication, etc.) that enables the recipient to retain a level of independent living in their home environment.

In an institutional setting the recipient is directly observed and interventions can be made swiftly. At home, the care recipient is likely to be observed only by each agency that visits the home; outside of this it may not be possible to make the required care intervention.
This is another compelling reason for the uptake of IoT technologies, whereby continuous monitoring of a care recipient can be employed.

This has two key benefits. 
First, personal data is retained within the home environment, and only exceptions are reported to care agencies or medical professionals. Second, since monitoring and analytics is performed locally, there is the opportunity to provide enhance interfaces that can inform the behaviour of the individual that is being monitored. This can support the more continuous provision of care between visits from external agencies.
\subsection{Emergency care}
\begin{figure}[!t]
\centering
\includegraphics[width=3.15in]{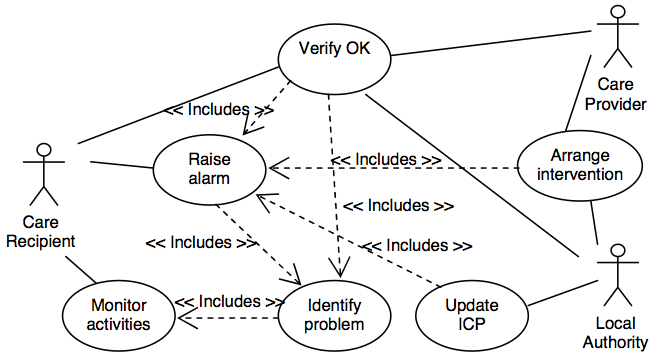}
\caption{Use case model for emergency care scenario.}
\label{fig:emergencycare}
\end{figure}

Emergencies within community health care settings are dealt with via conventional telephone based systems for all members of the UK public. Raising the alarm assumes that either the care recipient is conscious and able to reach either a telephone or an alarm system, or that a visitor is present who can raise the alarm. However, there are opportunities to improve the responsiveness of such a system. Increased monitoring can collate data to enable patterns in health conditions to be be recognised  and exceptions can be reported to relevant care agencies. In addition, informal carers can be brought into the system to assist during the period that the alarm has been raised and prior to the emergency services arriving on site. An important aspect here is that emergency interventions should be reflected in the ICP, thus providing a richer record of information for any subsequent care provision (Figure \ref{fig:emergencycare}).
%
%
%
\section{System design}
So far the actors in the system have been identified and their roles described. System interactions are described within the written use cases and use case models. The agent `lens' has enabled roles to be understood and specified in the context of the domain. The next step is to translate the business functionality in the system into microservices.
\subsection{Microservice characteristics}
Newman argues\cite{newman2015} that microservices are small autonomous services that work together, modelled around a business domain. This implies service autonomy, a key characteristic of agency, underlining the need for a service to own its own data model, or at least designed around a single responsibility principal\cite{bryant2016}.
As such, business functions that are related can be packaged in a similar fashion to a component, before being implemented as a microservice.


Johannes Th{\H o}nes advises that microservices\cite{thones2015} should be a small application that can be deployed, scaled and tested independently. To achieve this, a microservice needs to demonstrate resilience, flexibility and fault-tolerance, suggesting that method calls masquerading as services are too finely-grained.

From an architectural perspective microservices have come about to address emerging issues of scalability\cite{dragoni2017,dragoni,hasselbring2016}, with an emphasis upon reducing the overheads of messaging\cite{zeromq,pivotal,kubernetes}, containerisation\cite{docker} and microservice orchestration\cite{mesosphere,xu2008}. Many organisations are now realising that they can no longer design applications that will serve their business models going forward. Applications need to be able to be extended and augmented in the future, and their architecture should scale as required. Accordingly, it should be possible to retire functionality without breaking other services required by the business. As each cohesive microservice must retain its own data model, data replication is necessary to govern all of the data within an application, which creates an additional processing overhead\cite{viennot2015}.

\begin{table*}[htbp]
\centering
\caption{Assignment of tasks to microservices.}
\label{comparison}
\resizebox{\textwidth}{!}{%
\begin{tabular}{ |l|l|l| }
\hline
\textbf{Microservice}       & \textbf{Tasks}                                         \\
\hline\hline
Individual Care Plan Microservice            & 1. Create ICP\\
&2. Maintain ICP\\
&3. Record outcome of needs assessment\\
&4. Update record of care received\\
&5. Analytics on care received.                                                                                                                                           \\
\hline
Care Provider Microservice(s) &    1. Update care plan\\
 & 2. Deliver to care plan                                                                                                                                                                                                                                                                                                                                                            \\ \hline
Payment Processing Microservice  & 1. Manage payments for one provider\\
& 2. Manage payments where more than one provider is supplying the care                                                                                                                                             \\ \hline
Feedback Microservice      & Handles functionality for feedback from care recipient \\ \hline
\end{tabular}%
}
\end{table*}
For the community health care case study, Table \ref{comparison} illustrates the mapping of tasks to microservices.
\section{System architecture}
\begin{figure*}[htbp]
\centering
\includegraphics[width=\textwidth]{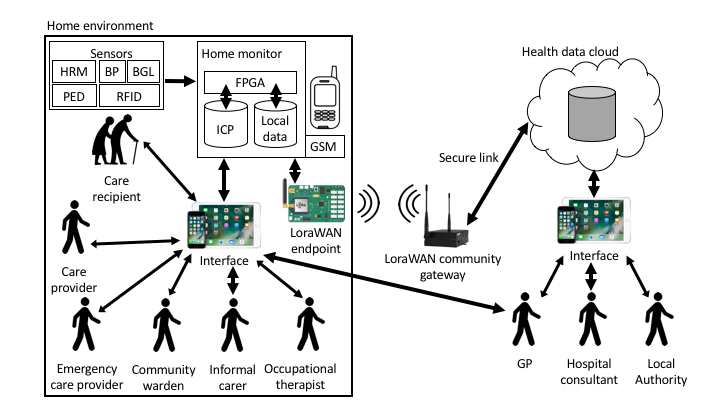}
\caption{System architecture for the Intelligent Community Alarm 2 (InCA2).}
\label{fig:inca2architecture}
\end{figure*}

Beer et al's Intelligent Community Alarm (InCA)\cite{beer2003} posited the compelling case to adopt a multiagent approach for the management of community healthcare services. InCA2 utilises IoT and microservices technologies to provide the system architecture as described in Figure \ref{fig:inca2architecture}. The InCA2 architecture makes use of elastic cloud resources to simplify the management of healthcare data. This resource is accessible by healthcare professionals and the local authority at either hospitals or a General Practitioner's (GP) surgery. A key departure from the original InCA architecture is the adoption of the following technologies within the care recipient's community environment:
\begin{itemize}
\item Wearable biosensors to provide monitoring of vital functions such as heart rate (HRM), blood pressure (BP), blood glucose level (BGL), activity (PED), as well as environment sensing such as movement/occupancy (PIR) and pressure pads within the home setting;
\item Reconfigurable embedded computational hardware within the care recipient's home, using Field Programmable Gate Arrays (FPGA) for the collection, processing and management of personal health data;
\item Wireless connectivity within the community using a Low Power Wide Area Network, in this case LoraWAN.
\end{itemize}
For the software architecture, a microservices approach has been adopted to provide future requirements scalability. As per the design of InCA, the preservation of personal health data is a primary concern for InCA2. Whilst personal data is being continuously generated through monitoring within the home environment, access to this is restricted based upon a) the role of the actor that is making a query, and b) the physical location of the actor.


%
\section{Experimentation}
We have identified two main challenges for the adoption of the INCA2 architecture. First, the need to ensure that raw data from the biosensors is managed in a way that minimises the quantity of data retained in the home, without compromising the quality of any subsequent decisions taken by healthcare professionals. Second, the adoption of low power WAN means that data transfer rates from the home environment to the health cloud are restricted to 50kbps. Therefore, we have developed a two-stage approach to the simulation of these challenges, whereby the Home Monitor unit divides its processing capabilities into two components. The first component allocates processing cycles to sensors at the edge of the network. The remaining component processes filtered data from the edge components, in conjunction with a local store of data and the Individual Care Plan, producing analytics for consumption by care professionals.
\subsection{Stage One: Sensor data filtering}
Stage one relates to the initial filtration of sensor data that is redundant. As per Algorithm 1, if the temperature of the care recipient is within a normal range then it is added to a buffer, which can be subsequently emptied at regular periods. If the temperature falls outside of this range, the data is sent onwards for further processing instantaneously.
\begin{algorithm}
\SetAlgoLined
\KwResult{Filter body temperature data.}
 \While{HomeMonitorActive}{
  \eIf{$ 36 < temp < 37$}{
   bufferDataStream\;
   }{
   sendData\;
  }
 }
 \caption{Filtering data from body temperature biosensor.}
\end{algorithm}
Algorithm 2 is an example of detecting exceptions such as a potential alarm condition. When the care recipient's heart rate is below 40 bpm and their blood pressure is below 90, an alarm condition is detected for the home monitor to take further action.
\begin{algorithm}
\SetAlgoLined
\KwResult{Forward alarm condition.}
 \While{HomeMonitorActive}{
  \eIf{$hrm < 40 || bp < 90 $}{
   sendAlarm\;
   }{
   //other exceptions\;
  }
 }
 \caption{Identify data exceptions such as an alarm condition.}
\end{algorithm}
\subsection{Stage Two: Home Monitor data analytics}
The second component receives filtered sensor data from the edge component, and combines this with a local data store to identify trends and patterns that may be of concern to care providers. To manage the data that is collected within the home, before subsequently transporting any reports to the health cloud via a LoraWAN community gateway, it is necessary to be able to manage the overall workload of the filtration and analytics. Table \ref{table:edgeparameters} illustrates the simulation parameters of the system to manage work in the Home Monitor unit, with the associated processing rule description in Algorithm 3.
\begin{table}
\centering
\caption{Managing the data filtration and analytics workload}
\begin{tabular}{|c|c|c|} \hline
\textbf{Name} & \textbf{Description} & \textbf{Value} \\ 
\hline
AnalyticsDeadline & Maximum time for processing buffer &0.5 sec \\
\hline
BufferStorage & Total edge item capacity & 100 \\
\hline
AlgorithmTime & Time to process one data item & 0.01 sec \\
\hline \end{tabular}
\label{table:edgeparameters}
\end{table}
\begin{algorithm}
\SetAlgoLined
\KwResult{Manage data processing workload.}
 \While{HomeMonitorSimulating}{
  addIncomingMsgToBuffer()\
  timer=startTimerThread()\\
  \For {each item in buffer}{
  executeAlgorithm(item)}
  
  \eIf{timer.time$ > $AnalyticsDeadline}{
   forwardBufferContents\;
   }{
   break\;
  }
 }
 \caption{Apportioning workload between Home Monitor edge components.}
\end{algorithm}
\section{Results}
The network simulation was performed using Omnet++ 5.1 on Linux Ubuntu 16.04.3, within Oracle Virtual Box 5.1.28. A dataset of sensor readings was used from the MIT Artificial Intelligence Lab (http://db.csail.mit.edu/labdata/labdata.html), to provide a suitable volume of monitoring data to be processed (approximately 100MB). The simulation duration was 1000 ticks.
For a given run, the filtering of data at the edge of the network resulted in a reduction of data traffic by 68\%, when compared with the total amount of sensor data being processed. Similar reductions in the time taken to process messages was also observed from 18.3 secs to 7.3 secs, suggesting a 59\% reduction in compute time. Perhaps more pertinent is the reduction in network bandwidth consumption of around 51\%, which places less demands upon the transmission of reports using the constrained LoraWAN gateway.
The introduction of workload management across the computational components of the Home Monitoring unit appears to provide its most significant benefit when the number of sensors is increased. The edge architecture is less susceptible to changes in the network, as the Home Monitoring unit shares processing between its components depending on what resource is available. The total bandwidth saving across the simulated network amounts to 57\% when incorporating workload management.
\section{Conclusions and future work}
%
%
%

%
This article describes the architecture of a community healthcare management system that utilises IoT and microservices technologies.

Our experiment via simulation of the scope for data collection and transfer demonstrates the efficacy of a two stage approach to sensor data filtration followed by subsequent processing, in order to reduce the transport of potentially redundant data. This enables LoraWAN networks to be utilised for non time critical reporting. Alarm conditions can be routed through the GSM network to relevant emergency services as required.

At present, the simulation applies two stages of rules to filter and process data from the sensor dataset. Additional processing such as the encoding of trends and analysis as graphs, will reduce the data transmission load on the LoraWAN gateway further. The Home Monitoring hardware has computational capacity that has not been fully exploited as yet.

The adoption of a microservices architecture has several benefits. First, it facilitates the design of services that directly support the fundamental use cases of care provision; services are more granular and therefore easier to assemble into composite service offerings. This helps create a system design that is responsive to emerging requirements in the future. Second, the cohesive workflow of microservices development leads towards greater software resilience as each service must fail without adversely affecting the overall system. Third, data privacy can be enforced by specific services that can negotiate and broker access over time between different actors (stakeholders).


There are three distinct areas of development for future work. First, to develop a reference architecture using traditional OO approaches, against which a microservice demonstrator can be compared in terms of classical software engineering metrics such as component reuse. Second, to perform a hardware-in-the-loop evaluation of the additional processing overhead required to support decentralised data replication across microservices, particularly when the number of care recipients increases considerably. Finally, to investigate the impact of data privacy protocols and policies, to enable marshalled analyses between the health care providers and the Local Authority.

\end{document}